\documentclass[12pt,a4paper]{article}
\usepackage[british]{babel}
\usepackage[dvips]{epsfig}
\usepackage{amssymb}
\title{
  {\vspace{-2cm} \normalsize
     \epsfig{figure=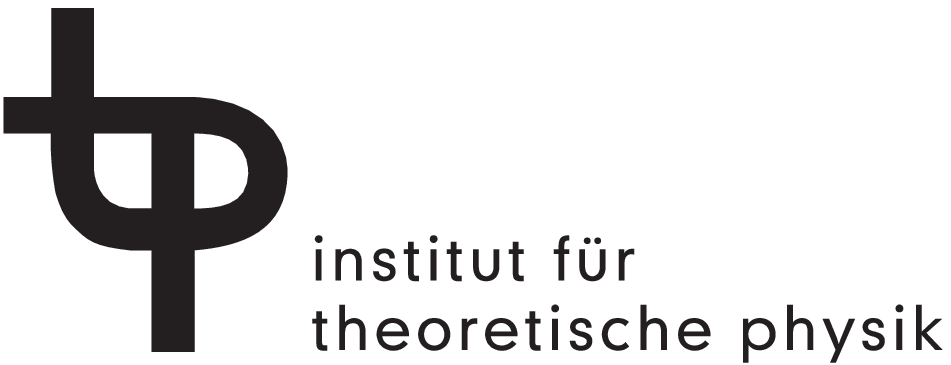,width=80mm}
     \hfill\parbox[b][30mm][t]{35mm}{MS-TP-04-5 \\
                                      hep-lat/0404004}  }\\[25mm]
The Volume Source Technique for flavor singlets:\\ a second look}
\author{F.~Farchioni, G.~M\"unster and R.~Peetz\\
        Institut f\"ur Theoretische Physik,
        Universit\"at M\"unster\\
        Wilhelm-Klemm-Str.~9, D-48149 M\"unster, Germany\\
        e-mail: farchion@uni-muenster.de}
\date{May 6, 2004}
\newcommand{\be}{\begin{equation}}
\newcommand{\ee}{\end{equation}}
\newcommand{\bea}{\begin{eqnarray}}
\newcommand{\eea}{\end{eqnarray}}

\newcommand{\Tr}{\mbox{Tr}}
\newcommand{\calQ}{{\mathit Q}}

\begin{document}
\maketitle

\begin{abstract} \normalsize
We reconsider the Volume Source Technique (VST) for the determination
of flavor singlet quantities on the lattice.  We point out a difficulty
arising in the case of fermions in real representations of the gauge
group and propose an improved version of the method (IVST) based on
random gauge transformations of the background configuration.  We
compare the performance of IVST with the method based on stochastic
estimators (SET).  We consider the case of the N=1 Supersymmetric
Yang-Mills Theory (SYM), where just one fermionic flavor is present,
the gluino in the adjoint representation, and only flavor singlet
states are possible.  The work is part of an inclusive analysis of the
spectrum of the lightest particles of the theory, based on the
simulation of the model on a $16^3\cdot32$ lattice with dynamical
gluinos in the Wilson scheme.
\end{abstract}
\newpage
\section{Introduction}
\label{sec1}

Supersymmetry (SUSY) is broken on the lattice owing to the finite
lattice spacing $a$.  We consider the N=1 Supersymmetric Yang-Mills
Theory (SYM) with gauge group SU(2) and Wilson discretization in the
fermion sector.  Here SUSY is also explicitly broken by the Wilson
term.  However, by properly tuning the (renormalized) gluino mass to
zero, SUSY is expected to be recovered in the continuum
limit~\cite{CuVe} with exponentially small $O(a)$ deviations.

Manifestation of SUSY occurs at the non-perturbative level, the most
interesting phenomenological implication being the expected ordering of
the bound-states of the theory in supermultiplets.  In the low energy
sector in particular, effective Lagrangians for SYM
predict~\cite{VeYa,FaGaSch} two Wess-Zumino supermultiplets.  The
spin-0 particles are represented by meson-like bound states of the
gluino and by glueballs, respectively, of opposite parity (this
classification is of course only valid in absence of mixings, which are
however expected).  The spin-$\frac{1}{2}$ particle of the multiplet is
in both cases a gluino-glue bound-state.

We focus here on the problem of determining  the masses of meson-like
gluino bound states.  Borrowing the terminology of QCD, these represent
``flavor singlet'' states.  Indeed, SYM resembles $N_f$=1 QCD, with the
quark in the fundamental representation replaced by the gluino in the
adjoint representation.  The lattice computation of flavor singlet
correlators is difficult because of the presence of disconnected
diagrams (see~\cite{Schilling} for a recent review on the topic).  The
exact evaluation of the correlator for these diagrams is not feasible
since it requires the trace over color and space-time indices of the
fermion propagator in the background of the gauge configuration, which
in turn involves the solution of an ``all-points to all-points''
inversion problem {\em for any given gauge configuration}. The first
approach to the subject was based on a volume source~\cite{VoSo}, the
so-called ``Volume Source Technique'' (VST).  For a given background
configuration the method delivers an estimate of the correlator which,
however, contains spurious terms represented by non-closed loops. 
In~\cite{VoSo}, where QCD was considered, it was argued that these
terms disappear in the ensemble-average on the basis of gauge
invariance.  In this paper we reconsider this argument more generally,
showing that it is not applicable to models where the fermions are in
real representations of the gauge group, as is the case for any
representation of SU(2) and for the adjoint representation of SU($N_c$).
We propose a new formulation of the method, based on random gauge
transformations of the background gauge configuration, which solves the
problem.  We shall refer to this in the following as to IVST (Improved
Volume Source Technique).  Due to the randomness introduced by the
gauge transformation, IVST is analogous to the well known
Stochastic Estimator Technique SET~\cite{SET}.  In both cases the
systematic error introduced by the computational procedure is converted
into a statistical one and can be controlled by increasing the number
of stochastic estimates. As a consequence IVST and SET can be directly
compared.

This work represents the sequel of a long-standing project having the
goal of a lattice verification of the non-perturbative low-energy
properties of SYM.  We refer to~\cite{Montvay} and the references
therein for the scope and goals of past studies.  The model is
simulated by means of the dynamical-gluino two-step multi-bosonic
algorithm. Details on the algorithm can be found in~\cite{CaetAl}.  The
present analysis is based on a sample of configurations of SU(2)~SYM on
a $16^3\cdot32$ lattice. Partial results have been reported
in~\cite{RoPe}.

In the next section we shall reconsider the theory of VST and propose
the improved version of it, IVST.  In section~3 the numerical results
will be presented, comparing IVST and SET; finally section~4 contains
our conclusions.
\section{The Volume Source Technique revisited}
\label{sec2}

In this section we consider lattice gauge theory with gauge group
SU($N_c$). The results of primary interest are for the gauge group SU(2)
or for models with fermions in the adjoint representation of the gauge
group. This includes SYM in particular. In the following, Greek letters
denote Dirac indices, Latin letters color, $\Tr_d$ and $\Tr_c$ are the
respective traces.  With the usual bilinears $\bar\psi(x)\Gamma\psi(x)$
as insertion operators for the singlet mesonic states, where $\Gamma=1$
or $\gamma_5$, the disconnected part of the mesonic correlator in the
background of a gauge configuration~$\{U\}$ can be written as
\be
C_{\Gamma,{\mbox{\scriptsize disc}} }[U](x_0-y_0) =
\frac{1}{V_s} \Tr_{d}[ \Gamma S(x_0)]\, \Tr_{d}[ \Gamma S(y_0)]\,,
\label{disc_corr}
\ee
where the time-slice sum $S(x_0)$ represents the trace over color and
space indices of the inverse fermion-matrix, i.e. the propagator in the
background of the gauge configuration $\{U\}$:
\be
 S_{\alpha\beta}(x_0) =
\sum_{\vec{x}} \Tr_{c} [\calQ^{-1}_{x \alpha, x \beta} ]\,.
\label{timeslice}
\ee
VST delivers an estimate of $S_{\alpha\beta}(x_0)$ at the price of a
single inversion for each value of the color and Dirac index.
The inversion problem with the volume-source $\omega_{V}$ reads
\be
\calQ Z = {\omega_{V}}^{[a, \alpha]}\,, \quad
(\omega_V^{[a, \alpha]})_{xb\beta}=\delta_{ab}\,\delta_{\alpha\beta}
\ee
with solution
\be
Z^{[a,\alpha]}_{xb\beta} =
[\calQ^{-1}\omega_{V}^{[a,\alpha]}]_{xb\beta} =
  \calQ^{-1}_{xb\beta,xa\alpha}
  + \sum_{y\neq x} \calQ^{-1}_{xb\beta,ya\alpha} \label{v_so_inv}\,.
\label{vol_sol}
\ee

When $Z^{[a,\alpha]}$ in the above equation is used to estimate
the time-slice sum~(\ref{timeslice}),
\be
S_{\alpha\beta}(x_0)\rightarrow \tilde{S}_{\alpha\beta}(x_0) =
\sum_{\vec{x},a} Z^{[a,\beta]}_{xa\alpha}\,,
\label{replace}
\ee
the last term in~(\ref{vol_sol}) yields contributions to the
disconnected part of the correlator~(\ref{disc_corr}) which represent
non-closed loops. Such elements of the inverse fermion-matrix with
$x\neq y$ are non-gauge-invariant and are canceled in the average over
the gauge-ensemble (which is gauge-invariant). However, there are also
contact terms in the correlator, which are potential sources of
systematic errors.

In the original work~\cite{VoSo}, which introduced VST in the context of
QCD, these unwanted terms were avoided by considering the
correlator
\be
\hat{C}_{\Gamma,{\mbox{\scriptsize disc}} }[U](x_0-y_0) =
\frac{1}{V_s} \Tr_{d}[ \Gamma \tilde{S}(x_0)]\, 
\Tr_{d}[ \Gamma \tilde{S}^{\dagger}(y_0)]
\label{dagger_corr}
\ee
with one of the time slices conjugated. Owing to the fact that the
product $\mathbf{3} \otimes \mathbf{3}$ of fundamental representations
of SU(3) does not contain the trivial representation, a gauge invariant
contact term does not appear. The argument holds more generally for the
fundamental representation of SU($N_c$) for $N_c>2$.

In the case of gauge group SU(2), which has real representations only,
or in the case of the adjoint representation of SU($N_c$), this
prescription, however, does not help. For SU(2) the product of two
fundamental representations contains the trivial one, which leads to
non-vanishing contact terms again. The same is true for the adjoint
representations of SU($N_c$).

We now want to consider the gauge invariance of the contact terms in
detail. We focus on the correlator (\ref{disc_corr}), for
(\ref{dagger_corr}) the discussion is analogous.

Consider the following average over gauge transformations $g(x)$ ({\em
gauge-average}):
\be
\langle\tilde{S}_{\alpha\beta}(x_0)\tilde{S}_{\gamma\delta}(y_0)
\rangle_g =
\left\langle \sum_{\vec{x},w,a} Q^{-1}_{xa\alpha,wa\beta} [U^g]\,
             \sum_{\vec{y},z,b} Q^{-1}_{yb\gamma,zb\delta}[U^{g}]
\right\rangle_g.
\label{CrossTerm}
\ee
The gauge-average induces an average over the gauge-orbit $\{U^g\}$.
Using
\be
Q_{x,y}^{-1} [U^g]= g^\dagger(x) Q_{x,y}^{-1} [U] g(y)
\label{trans_prop}
\ee
and the general formula
\be
\langle g_{ab}(x) g^{-1}_{a'b'}(x')\rangle_g =
A\,\delta_{xx'}\delta_{ab'}\delta_{a'b}\,,
\quad
A=\left\{ \begin{array}{ll}  \frac{1}{N_c},      & \mbox{fundamental}
                \\           \frac{1}{N_c^2-1},  & \mbox{adjoint}
            \end{array}
  \right.
\label{gauge_av_form}
\ee
(in the adjoint representation $g$ are real orthogonal matrices of
dimension $N_c^2-1$), the gauge-average of~(\ref{CrossTerm}) reads for
$x_0\neq y_0$ 
\be
\langle\tilde{S}_{\alpha\beta}(x_0)\tilde{S}_{\gamma\delta}(y_0)
\rangle_g =
\sum_{\vec{x}} \Tr_c[Q^{-1}_{x\alpha,x\beta}]
\sum_{\vec{y}}\Tr_c[Q^{-1}_{y\gamma,y\delta}] +
A\,\sum_{\vec{x},\vec{y}}\Tr_c[Q^{-1}_{x\alpha,y\beta}
Q^{-1}_{y\gamma,x\delta}]\,.
\label{gauge_av}
\ee
The above expression represents the {\em gauge-invariant part} of
$\tilde{S}_{\alpha\beta}(x_0)\tilde{S}_{\gamma\delta}(y_0)$.

Let us now consider the {\em ensemble-average} of
$\tilde{S}_{\alpha\beta}(x_0)\tilde{S}_{\gamma\delta}(y_0)$.  In the
limit of infinite statistics any given gauge-orbit is completely
covered, implying that the ensemble-average delivers in particular a
gauge-average.  Using the result in~(\ref{gauge_av}) this implies
\be
\left\langle\tilde{S}_{\alpha\beta}(x_0)\tilde{S}_{\gamma\delta}(y_0)
\right\rangle_U =
\left\langle S_{\alpha\beta}(x_0)S_{\gamma\delta}(y_0)\right\rangle_U
+ A \left\langle\sum_{\vec{x},\vec{y}}
\Tr_c[Q^{-1}_{x\alpha,y\beta} Q^{-1}_{y\gamma,x\delta}]
\right\rangle_U.
\ee
We thus obtain that replacement~(\ref{replace}) in~(\ref{disc_corr})
produces an error-term for the {\em full} disconnected correlator
\bea
\tilde{C}_{\Gamma,{\mbox{\scriptsize disc}} }(x_0-y_0) &=&
C_{\Gamma,{\mbox{\scriptsize disc}} }(x_0-y_0) +
\Delta C_{\Gamma,{\mbox{\scriptsize disc}} }(x_0-y_0)\,,
\label{disc_corr_full}
\\
\Delta C_{\Gamma,{\mbox{\scriptsize disc}} }(x_0-y_0) &=&
A\,\frac{1}{V_s} \left\langle\sum_{\vec{x},\vec{y}}
\Tr_{c}[\Tr_d[Q^{-1}_{x,y}\Gamma]\,
\Tr_d[Q^{-1}_{y,x}\Gamma]]\right\rangle_U.
\label{err_corr}
\eea
The conclusion is that the error-term in~(\ref{vol_sol}) produces a
systematic error in the correlator, {\em which does not vanish in the
ensemble-average even in the limit of infinite statistics}.  This error
is due to gauge invariant contact terms in the correlator, as shown
above. The spurious term resembles the connected contribution
\be
C_{\Gamma,{\mbox{\scriptsize conn}} }[U](x_0-y_0) =
-f\,\frac{1}{V_s} \sum_{\vec{x},\vec{y}}\Tr_{cd}[Q^{-1}_{x,y}\Gamma\,
Q^{-1}_{y,x}\Gamma]\,,\quad
f=\left\{
    \begin{array}{ll}
1,     & \mbox{fundamental}
\\
2,    & \mbox{adjoint}
    \end{array}
 \right.
\label{conn_corr}
\ee
the only difference being in the Dirac structure and the numerical
factor. This outcome is not surprising considering that gauge invariance
strongly constrains the space-time and color structure. We have checked
the presence of the error-term numerically for both types of correlators
(\ref{disc_corr}) and (\ref{dagger_corr}) for gauge group SU(2), see
section~\ref{sec2.1}.

At this point we make the simple observation that the error is removed
by using the gauge-average of $\tilde{S}_{\alpha\beta}(x_0)$ to
determine the time-slice sums, since
\be
\left\langle \tilde{S}_{\alpha\beta}(x_0)\right\rangle_g =
S_{\alpha\beta}(x_0)\,.
\label{avg_ts}
\ee
In practice this is obtained by averaging 
$\tilde{S}_{\alpha\beta}(x_0)$ over a sufficiently large number $N_g$ of
gauge configurations obtained from the original one by random gauge
transformations~\cite{RoPe}\footnote{After the completion of this study we
noticed that the use of random gauge transformations in VST was recently
pointed out in~\cite{Schilling}.} $g(x)$, namely with a flat probability
distribution
\be
\frac{dp}{dg}=1\,,
\ee
where $dg$ denotes the Haar measure on the gauge group.  Besides
solving the problem of the error~(\ref{err_corr}) in the correlator,
the method brings the additional benefit of disentangling the
systematic error inherent in VST from the statistical one: in the limit
of an infinite number of random gauge transformations
$N_g\rightarrow\infty$ the former goes to zero, only the second one
surviving.  In this view the improved version of VST (IVST in the
following) is analogous to the techniques based on stochastic
estimators (SET), the randomness of the source being replaced by that
of the gauge transformation.\footnote{Actually on the basis of
(\ref{trans_prop}) IVST could be seen as a stochastic estimator method
with a particular stochastic volume source.} This allows for a direct
comparison of the two methods, which is carried out in the next
section.
\section{Numerical analysis}
\label{sec2.1}

The simulation parameters of the gauge sample are $\beta=2.3$ and
$\kappa=0.194$. The estimated value of the lattice spacing is, in QCD
units, $a\approx 0.06$~fm ($a^{-1}\approx$ 3.3~GeV); there are
indications~\cite{FaPe} that the gluino is still relatively heavy
($m_{\tilde{g}}\gtrsim 200$~MeV on the basis of QCD-inspired
arguments). The set-up of the two-step multi-bosonic algorithm
algorithm is the same as in~\cite{FaetAl}, and $\sim$4000 thermalized
configurations were stored every 5 or 10 cycles. In order to obtain an
estimate of the autocorrelation time of the disconnected part of the
mesonic correlator, an analysis of the autocorrelation time of the
smallest eigenvalue of the hermitian fermion-matrix was performed.  The
procedure is based on the expectation that the disconnected part of the
mesonic correlator is strongly related to the infrared behavior of the
fermion-matrix.  After that, a subsample of 218 supposedly uncorrelated
configurations was selected.  This constitutes the sample for the
numerical analysis.
\subsection{Time-slice sums}

For each configuration, 50 estimates of the time-slice
sums~(\ref{timeslice}) were performed, each obtained by applying a
random gauge transformation on the original gauge configuration as
explained in the previous section.  The computations were performed in
64 bit arithmetic. Improved summation techniques were employed to
ensure accuracy.

In the case of SYM the Majorana nature of the gluino field (invariance
under charge conjugation) allows to compute the inverse of the
fermion-matrix for only half of the matrix-elements in Dirac space.
This implies that, in the case of SU(2) SYM, only 6 fermion-matrix
inversions must be performed for each configuration, compared to 12
inversions needed for QCD.  So the total number of inversions
$N_{\textit{\scriptsize inv}}$ required for a determination of the
time-slice sum with $N_{\textit{\scriptsize est}}$ estimates is
$N_{\textit{\scriptsize inv}}=6N_{\textit{\scriptsize est}}$.%
\footnote{$N_{\textit{\scriptsize est}}$ coincides with $N_g$ of
previous section.  The change of notation is for the sake of the
homogeneity when comparing with SET.}

As IVST is based on stochastic estimations, a comparison with
stochastic-source methods SET suggests itself. We consider the SET
variant with complex $\mathbf{Z}_2$ noise in the spin explicit variant
SEM~\cite{SEM}.  In this case each estimate of the time-slice sum is
obtained by inverting the fermion-matrix with source
$
(\omega_S^{[\alpha]})_{xb\beta} =
\delta_{\alpha\beta}\,\eta^{[\alpha]}_{xb}
$
where $\eta^{[\alpha]}_{xb}$ are independent stochastic variables chosen
at random among $\frac{1}{\sqrt{2}}(\pm 1 \pm i)$.  For SET one has then
$N_{\textit{\scriptsize inv}}=2N_{\textit{\scriptsize est}}$.  (Again a
factor of two less comes from the symmetry of SYM.)  We computed 165
estimates of the time-slice sums, in this case using 32-bit arithmetic.

\begin{figure}
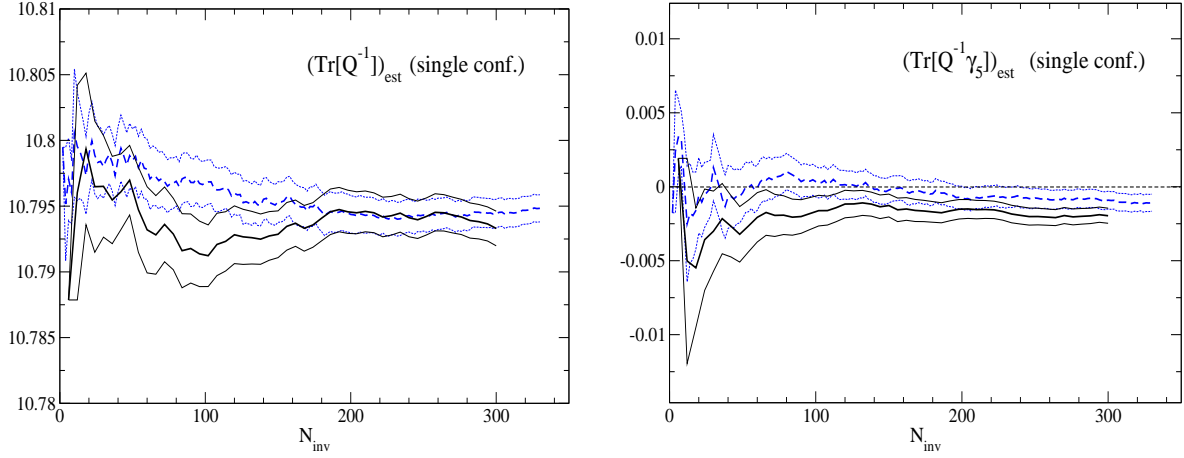

\hspace{-0.5cm}
\parbox{8cm}{
  \centerline
       {\epsfig{file=single_config_trQ.eps, width=7.5cm,height=6cm}}}
\parbox{8cm}{
  \centerline
       {\epsfig{file=single_config_trg5Q.eps, width=7.5cm,height=6cm}}}
\caption{Evolution of the estimated value of $\Tr[Q^{-1}]$ and
         $\Tr[\calQ^{-1}\gamma_5]$ for a chosen configuration as a
         function of the number of the needed inversions (with
         error-bounds). 
         Full lines: IVST, dashed lines: SET.
\label{fig:trQg_single}}
\end{figure}

\begin{figure}
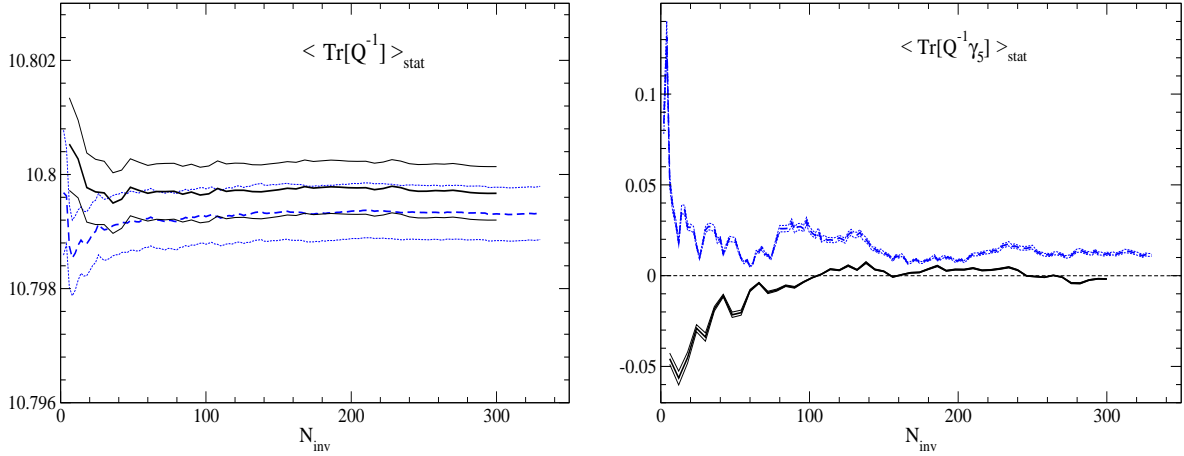

\hspace{-0.5cm}
\parbox{8cm}{
  \centerline
       {\epsfig{file=ensemble_trQ.eps, width=7.5cm,height=6cm}}}
\parbox{8cm}{
  \centerline
       {\epsfig{file=ensemble_trg5Q.eps, width=7.5cm,height=6cm}}}
\caption{Evolution of the average value of $\Tr[Q^{-1}]$ and
         $\Tr[\calQ^{-1}\gamma_5]$ over the complete sample as a
         function of the number of the needed inversions (with
         error-bounds).
         Full lines: IVST, dashed lines: SET.
\label{fig:trQg_ensemble}}
\end{figure}

In Fig.~\ref{fig:trQg_single} the evolution of the estimated value of
$\Tr[Q^{-1}\Gamma]\equiv\sum_{x_0}\Tr_d[S(x_0)\Gamma]$ for a chosen
configuration is displayed as a function of the number of needed
inversions $N_{\textit{\scriptsize inv}}$.  The error bounds represent
the statistical uncertainty on the stochastic estimation.  For both
IVST and SET the value stabilizes after 150-200 inversions, with
compatible results.  This test on a single configuration only serves as
a cross-check of the two methods, the physical information being
contained in the ensemble-averages, Fig.~\ref{fig:trQg_ensemble}.  In
the scalar case the two methods give compatible results after only 50
inversions. In the pseudoscalar case, fluctuations much larger than the
error-bounds indicate additional effects.  The fluctuations appear to
be more relevant for SET, where 32-bit arithmetic was used. Moreover,
in the latter case the estimate has an offset, while in the case of
IVST the expected value (zero) is approached after $\sim$100
inversions.

\begin{figure}
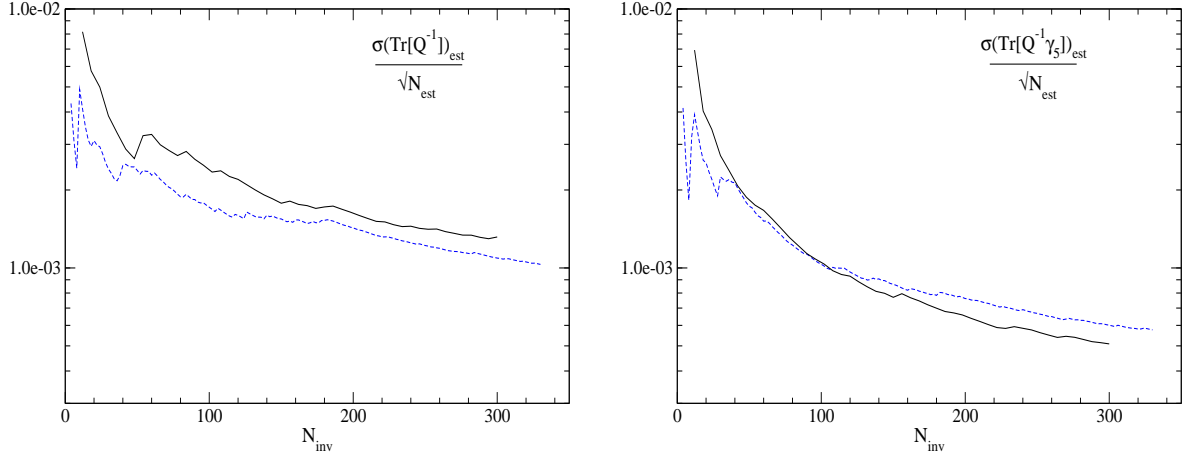

\hspace{-0.5cm}
\parbox{8cm}{
  \centerline
       {\epsfig{file=conf_gI_error.eps, width=7.5cm,height=6cm}}}
\parbox{8cm}{
  \centerline
       {\epsfig{file=conf_g5_error.eps, width=7.5cm,height=6cm}}}
\caption{Evolution of the statistical error of the estimated value of
         $\Tr[Q^{-1}]$ and $\Tr[\calQ^{-1}\gamma_5]$ for the same
         configuration as in Fig.~1, as a function of the number of the
         needed inversions.
         Full lines: IVST, dashed lines: SET.
\label{fig:trQgErr_single}}
\end{figure}

\begin{figure}
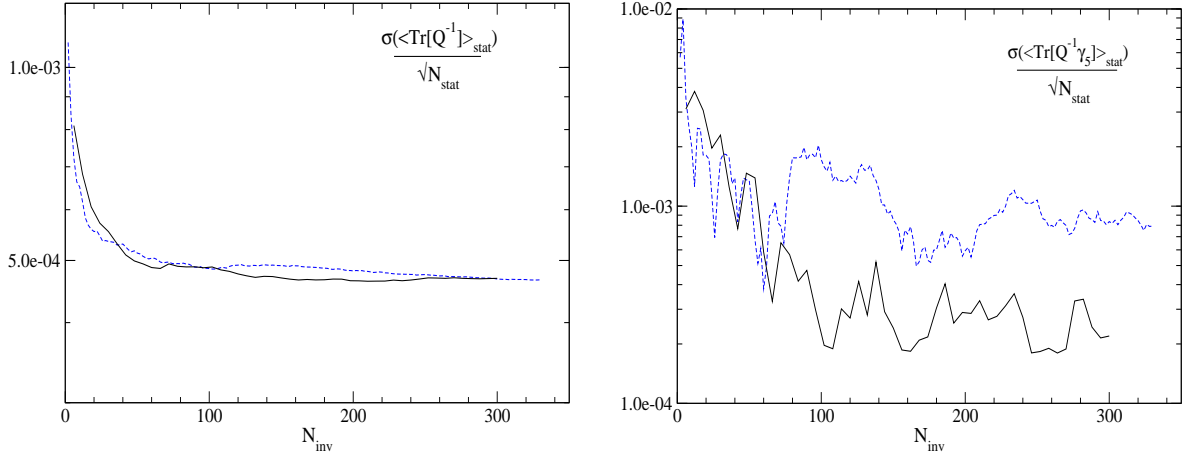

\hspace{-0.5cm}
\parbox{8cm}{
  \centerline
       {\epsfig{file=ensemble_gI_error.eps, width=7.5cm,height=6cm}}}
\parbox{8cm}{
  \centerline
       {\epsfig{file=ensemble_g5_error.eps, width=7.5cm,height=6cm}}}
\caption{Evolution of the statistical error on the average value of
         $\Tr[Q^{-1}]$ and $\Tr[\calQ^{-1}\gamma_5]$ over the complete
         sample as a function of the number of the needed inversions.
         Full lines: IVST, dashed lines: SET.
\label{fig:trQgErr_ensemble}}
\end{figure}

The evolution of the statistical error of the estimation for one
configuration is displayed in Fig.~\ref{fig:trQgErr_single}, showing the
a priori non-obvious result that the two methods introduce the same
amount of stochastic uncertainty.  The error in the estimation of the
ensemble-average is shown in Fig.~\ref{fig:trQgErr_ensemble}.  We see
that in both cases the error stabilizes after 100 inversions.  In the
pseudoscalar case, IVST seems to out-perform SET, although the large
instabilities prevent us from drawing firm conclusions.
\subsection{Correlators and masses}

\begin{figure}
\hspace{4cm}
\parbox{8cm}{
  \centerline
       {\epsfig{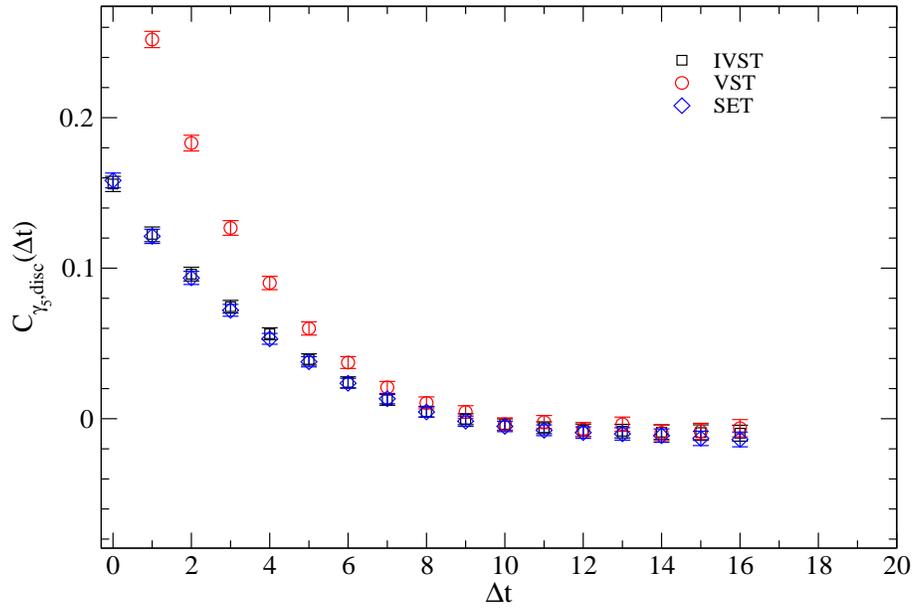}}}
\caption{The disconnected pseudoscalar correlator 
         $C_{\Gamma,{\mbox{\scriptsize disc}} }(\Delta t)$.
\label{fig:corr}}
\end{figure}

\begin{figure}
\hspace{4cm}
\parbox{8cm}{
  \centerline
       {\epsfig{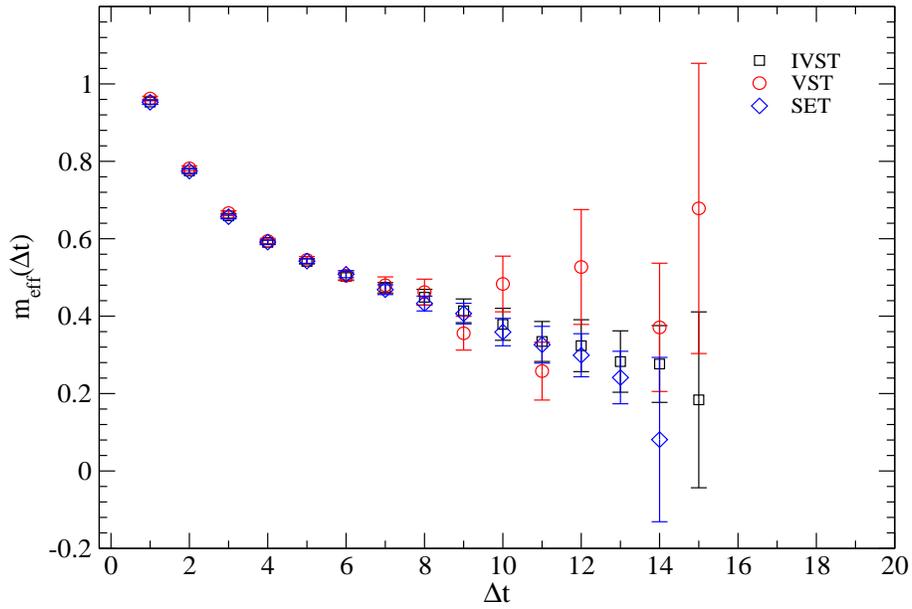}}}
\caption{The effective mass of the pseudoscalar meson.
\label{fig:mass}}
\end{figure}

In order to show the effect of the error-term~(\ref{err_corr}), we
computed the disconnected correlator in two ways: i) following the
correct procedure according to Eq.~(\ref{avg_ts}) (IVST); ii)
performing the gauge-average as in~(\ref{gauge_av}). As one can see in
Fig.~\ref{fig:corr} for the pseudoscalar meson, the error-term produces
a sizeable effect on the disconnected correlator.  IVST and SET are in
good agreement.  The effective mass is shown in Fig.~\ref{fig:mass}.
The impact of the error on the effective mass is suppressed in the
first time-slices where the connected contribution~(\ref{conn_corr})
plays a larger role.  However in the last time-slices, where the
disconnected contribution dominates, the effect of the error-term
shows-up in the form of a pronounced instability of the effective mass
as a function of the time-separation (for $\Delta t$=13 an estimate is
not even possible).  In the last few time-separations $\Delta t=14,15$,
IVST delivers a better result compared to SET (no estimate is possible
with SET for $\Delta t=15$). Since the disconnected contribution to the
mesonic correlator is essentially of infrared nature, the region of
large time-separations is important for the determination of masses.
\section{Conclusions}
\label{sec5}

We propose an improved version of the Volume Source Technique (IVST)
which eliminates erroneous contact terms in the case of fermions in real
representations of the gauge group. The improved version is based
on random gauge transformations and is analogous to stochastic
estimator methods (SET).  Comparison between IVST and SET shows
agreement and substantial equivalence.  In few cases, e.g.\ for the
determination of effective masses, IVST seems to give slightly better
results. A study with higher statistical precision should put these
observations on firmer ground.
\vspace*{1em}

\noindent
{\large\bf Acknowledgement}

\noindent
The computations were performed on the Cray T3E and JUMP systems at NIC
J\"ulich, the PC clusters at the ZIV of the University of M\"unster
and the Sun Fire SMP-Cluster at the Rechenzentrum of the RWTH Aachen.


\newpage

\begin{thebibliography}{99}
%
\bibitem{CuVe}
G.~Curci and G.~Veneziano,
Nucl.\ Phys.\ {\bf B\,292} (1987) 555.
%
\bibitem{VeYa}
G.~Veneziano and S.~Yankielowicz,
Phys.\ Lett.\ {\bf B\,113} (1982) 231.
%
\bibitem{FaGaSch}
G.R.~Farrar, G.~Gabadadze and M.~Schwetz,
Phys.\ Rev.\ {\bf D\,58} (1998) 015009.
%
\bibitem{Schilling}
K.~Schilling, H.~Neff and T.~Lippert, \texttt{hep-lat/0401005}.
%
\bibitem{VoSo}
Y.~Kuramashi, M.~Fukugita, H.~Mino, M.~Okawa and A.~Ukawa,
Phys.\ Rev.\ Lett.\ {\bf 72} (1994) 3448.
%
\bibitem{SET}
S.~J.~Dong and K.~F.~Liu,
Nucl.\ Phys.\ Proc.\ Suppl.\ {\bf 26} (1992) 353;
Phys.\ Lett.\ {\bf B\,328} (1994) 130.
%
\bibitem{Montvay}
I.~Montvay,
Int.\ J.\ Mod.\ Phys.\ {\bf A\,17} (2002) 2377.
%
\bibitem{CaetAl}
I.~Campos, A.~Feo, R.~Kirchner, S.~Luckmann, I.~Montvay, G.~M\"unster,
K.~Spanderen and J.~Westphalen,
Eur.\ Phys.\ J.\ {\bf C\,11} (1999) 507.
%
\bibitem{RoPe}
R.~Peetz, Ph.~D.\ Thesis, University of M\"unster, 2003.
%
\bibitem{FaPe} F.~Farchioni and R.~Peetz, in preparation.
%
\bibitem{FaetAl}
F.~Farchioni, A.~Feo, T.~Galla, C.~Gebert, R.~Kirchner, I.~Montvay,
G.~M\"unster and A.~Vladikas,
Eur.\ Phys.\ J.\ {\bf C\,23} (2002) 719.
%
\bibitem{SEM}
J.~Viehoff {\it et al.} [SESAM Collaboration],
Nucl.\ Phys.\ Proc.\ Suppl.\ {\bf 63} (1998) 269.
%
\end{thebibliography}
\end{document}